\documentclass[prd,showpacs,twocolumn,superscriptaddress,floatfix,nofootinbib,10pt]{revtex4-2}
\usepackage{setspace}
\usepackage[utf8]{inputenc}
\usepackage{float}
\usepackage{amsmath,amssymb,amsfonts,bm}
\usepackage{graphicx}
\usepackage{multirow}
\usepackage[usenames,dvipsnames]{color}
\allowdisplaybreaks
\usepackage[
colorlinks=true,
linkcolor=blue,
breaklinks=true,
urlcolor=blue,
citecolor=blue]{hyperref}

\usepackage{orcidlink}
\usepackage{subfigure}
\usepackage{tikz-feynman}
\usepackage{amsmath,cases}
\usepackage{xcolor}
\usepackage{physics}
\usepackage{booktabs}
\usepackage{ulem}
\usepackage{soul}
\usepackage[thicklines]{cancel}
\setstcolor{green}

\definecolor{green}{rgb}{0,0.6,0}

\newcommand{\mev}{\textrm{ MeV}}

\newcommand{\BeiHang}{\affiliation{School of Physics, Beihang University, Beijing, 102206, China}}

\newcommand{\IFIC}{\affiliation{Departamento de F\'{\i}sica Te\'orica and IFIC, Centro Mixto Universidad de
Valencia-CSIC Institutos de Investigaci\'on de Paterna, Apartado 22085,
46071 Valencia, Spain}}

\newcommand{\GXNU}{\affiliation{Department of Physics, Guangxi Normal University, Guilin 541004, China}}

\newcommand{\GXZD}{\affiliation{Guangxi Key Laboratory of Nuclear Physics and Technology, Guangxi Normal University, Guilin 541004, China}}

\newcommand{\CSU}{\affiliation{School of Physics, Central South University, Changsha 410083, China}}

\newcommand{\ITP}{\affiliation{CAS Key Laboratory of Theoretical Physics, Institute of Theoretical
Physics, Chinese Academy of Sciences, Beijing 100190, China}}

\newcommand{\FPU}{\affiliation{Departamento de F\'{\i}sica,  Universidade Federal do Piau\'{\i}, 64049-550 Teresina, Piau\'{\i}, Brasil}}

\begin{document}
\title{Testing the molecular nature of the $\Omega(2012)$ with the $\psi(3770) \to \bar{\Omega} \bar{K} \Xi$ and $\psi(3770) \to \bar{\Omega} \bar{K} \Xi^*(1530) (\bar{\Omega} \bar{K} \pi \Xi)$ reactions}

\begin{abstract}
	We report on the reactions  $\psi(3770)\to \bar{\Omega}^+ \bar{K} \Xi $ and $\psi(3770)\to \bar{\Omega}^+ \bar{K}\Xi^*(1530) \;(\Xi^*(1530)\to \pi\Xi$), and calculate the mass distributions  $\frac{{\rm d}\Gamma}{{\rm d}M_\text{inv}(\bar{K}\Xi)}$ and $\frac{{\rm d}\Gamma}{{\rm d}M_\text{inv}(\bar{K}\Xi^*)}$, respectively. We obtain clear peaks corresponding to the $\Omega(2012)$. From the decay of $\psi(3770)\to \bar{\Omega}^+ \bar{K}\Xi^*$, we also get a second, broader, peak around $2035\,\rm MeV$, which comes from the $\Omega(2012)$ decay to $\bar{K}\Xi^*$. This second peak is closely tied to the molecular picture of the $\Omega(2012)$ with the  $\bar{K}\Xi^*(1530)$ and $\eta\Omega$ components. Its observation would provide a boost to the molecular picture of the $\Omega(2012)$.
\end{abstract}

\author{Jing Song}
\BeiHang%
\IFIC%
%
\author{Wei-Hong Liang\orcidlink{0000-0001-5847-2498}}%
\email[Corresponding author: ]{liangwh@gxnu.edu.cn}
\GXNU%
\GXZD%
%
\author{Chu-Wen Xiao \orcidlink{0000-0001-5303-8350}}
\email[]{xiaochw@gxnu.edu.cn}
\GXNU%
\GXZD%
\CSU%
%
\author{Jorgivan Morais Dias \orcidlink{0000-0002-0354-4711}}%
\email[]{jorgivan.mdias@itp.ac.cn}
\ITP%
\FPU%
%
\author{Eulogio Oset \orcidlink{0000-0002-4462-7919}}%
\email[]{Oset@ific.uv.es}
\GXNU%
\IFIC%

\maketitle

\section{Introduction}\label{sec:Intr}

The $\Omega(2012)$ state discovered by the Belle Collaboration~\cite{Belle:2018mqs} offers one more example of a successful theoretical prediction come true. 
Indeed, in Refs.~\cite{Hofmann:2006qx,Sarkar:2004jh} it was predicted as a dynamically generated state from the interaction of the octet of pseudoscalar mesons with the baryons of the $\Delta(1232)$ decuplet. 
In particular, it would correspond to a molecular state of the $\bar{K} \Xi^*(1530)$ and $\eta \Omega$. 
Since the state was observed in the $\bar{K} \Xi$ invariant mass distribution, a decay channel of $\bar{K} \Xi^*(1530)$, papers claiming the molecular nature also incorporate the $\bar{K}\Xi$ channel~\cite{Valderrama:2018bmv,Lin:2018nqd,Pavao:2018xub,Huang:2018wth,Lu:2020ste,Ikeno:2020vqv,Ikeno:2022jpe,Liu:2020yen,Zeng:2020och}. 
As usual, there are alternative proposals, concretely as a $P$-wave excited $3/2^-$ state of the quark model~\cite{Xiao:2018pwe,Aliev:2018yjo,Aliev:2018syi,Polyakov:2018mow,Liu:2019wdr,Arifi:2022ntc,Wang:2022zja}, but other quark models, allowing for the formation of hadron pairs, also claim a molecular nature of the state~\cite{Wang:2007bf,BESIII:2023mlv}. 
Application of the Weinberg compositeness condition, as done in Ref.~\cite{Gutsche:2019eoh}, also leads to the conclusion of a molecular state. 

After the discovery, tests were conducted to support or refute the molecular nature of the $\Omega(2012)$ state, 
and in Ref.~\cite{Belle:2019zco} an experiment was done looking at the $\bar{K} \pi \Xi$ decay mode, 
which would come from the $\Xi^*(1530)$ decay into $\pi \Xi$ of the $\bar{K} \Xi^*(1530)$ molecular component.
A small fraction with an upper threshold of $11.9\%$ relative to the $\bar K \Xi$ decay mode was reported, just in the limit which could disqualify the molecular pictures \cite{Lu:2020ste,Ikeno:2020vqv,Ikeno:2022jpe}, 
but this was reanalyzed later with the result \cite{Belle:2022mrg}:
$R^{\Xi\pi \bar{K}}_{\Xi \bar{K}} = 0.99 \pm 0.26 \pm 0.06$, 
in line with the predictions of Ref.~\cite{Pavao:2018xub}, giving support for the molecular picture. 

In the present work, we propose a reaction, amenable of implementation at BESIII, to look for the $\Omega(2012)$ and again, 
making tests of the molecular nature of the state. 
The reaction  is $\psi(3770) \to \bar{ \Omega}^+ \Omega(2012)^-$, and we propose to look at the two decay modes of the $\Omega(2012)^-$, namely, the $\bar{K}\Xi$ and $\bar{K} \Xi^*(1530)$ ($\bar{K} \pi \Xi$). 
The idea is inspired in the recent BESIII work in which they studied the $\psi(3686)$ decay to $\bar{\Xi}^ +K^- \Lambda$, 
claiming to see the $\Xi(1820)$ state but with a width substantially larger than the one reported in the  Particle Data Group (PDG)~\cite{ParticleDataGroup:2024cfk}. 
This was interpreted in Ref.~\cite{Molina:2023uko} as an indication that the experiment observed the two $\Xi(1820)$ states previously predicted in Ref.~\cite{Sarkar:2004jh}.
Subsequently, a proposal of experiment was made using the $\Omega_c\to \pi^+ (\pi^0, \eta) \pi \Xi^*, \pi^+ (\pi^0, \eta) K \Sigma^*$ reactions~\cite{Liang:2024fsv}. 
But closer to the work proposed here, another reaction feasible at BESIII was suggested using the $\psi(3686)\to \bar{\Xi}^+\bar{K}^0 \Sigma^{*-}(\pi^-\Lambda )$ decay~\cite{Duan:2024ygq}.  
The $\bar{K}^0 \Sigma ^{*-}$ is one of the building blocks of the $\Xi(1820)$ resonances, 
and the second $\Xi$ state at $1875 \mev$ is close to the $\bar{K}^0 \Sigma ^{*-}$ threshold.
By producing this state, which decays to $\bar{K}^0\Sigma^{*-}$, with further decay of the $\Sigma ^{*-}$ to $\pi ^- \Lambda$, one succeeds in suppressing the production of the first $\Xi$ state at $1820 \mev$, giving more chances to see the second state at higher energy. 
This reaction would then be complementary to the one of Ref.~\cite{BESIII:2023mlv}, where the excitation of the $\Xi$ state of lower energy was prominent. 

In the present reaction, we combine the ideas of Ref.~\cite{Molina:2023uko} and Ref.~\cite{Duan:2024ygq} to look for the $\Omega(2012)$ state, by studying the $\psi(3770) \to \bar {\Omega}^+ \Omega(2012)^-$ reaction and looking for two final states, which can show the clean signal of the $\Omega(2012)$.  
Similarly to what was done in the BESIII experiment with the $\Xi(1820)$ decaying into the $\bar{K} \Lambda$, we look now at the $\Omega(2012)$ decaying to $\bar{K} \Xi$, and similarly to what is proposed in Ref.~\cite{Duan:2024ygq} looking at the $\bar{K} \Lambda \pi$ decay of the $\Xi(1820)$, we look now into the decay of the $\Omega(2012)$ into $\bar{K} \pi \Xi$.  
These are also the two decay channels of the $\Omega(2012)$ that were considered in the Belle experiments~\cite{Belle:2018mqs,Belle:2019zco,Belle:2022mrg}. 
In this way we have the chance to show the production of the $\Omega(2012)$ in these reactions and get a measure of the $\bar{K} \Xi^*(1530)$ component of the $\Omega(2012)$. 

\section{formalism}\label{sec:Formalism}

In order to describe the $\bar{K}\Xi$ and $\bar{K}\Xi^*$  mass distributions in the  $\psi(3770)\to \bar{\Omega}^+\Omega^- (2012)$ decay, we will employ two mechanisms to be discussed next. 

\subsection{$\psi(3770)\to \bar{\Omega}^+ \bar{K} \Xi $}

The first thing we must look at is the $\psi(3770)$ decay into $\bar{\Omega}^+$ and the $\bar{K}\Xi^*$ and $\eta\Omega$ components of the $\Omega(2012)$. 
This we do, following Ref.~\cite{Duan:2024ygq}, by considering that the $\psi(3770)$ is a $c\bar{c}$ state and, thus, a singlet of SU(3). 
We then use the SU(3) Clebsch-Gordan coefficients of $8 \otimes 10\to10$, obtaining the coefficients in Table~\ref{tableI}, with the intrinsic phase conventions of the isospin doublets ($\bar{K}^0, \;-K^-$), ($\Xi^{*0}, \;\Xi^{*-}$).
\begin{table}[!b]
	\centering
	\caption{SU(3) Clebsch-Gordan coefficients for the relevant $8 \otimes 10\to10$ transitions.}
	\label{tableI}
	\setlength{\tabcolsep}{10pt}
	\begin{tabular}{c|ccc}
	\toprule
	C-G coefficients & $\bar{K}^0 \Xi^{*-} $ & ${K}^- \Xi^{*0} $ &  $\eta\Omega$\\
	\midrule
	&$\dfrac{1}{2}$ & $\dfrac{1}{2}$ & $\dfrac{1}{\sqrt{2}}$ \\
	\bottomrule
	\end{tabular}
\end{table}

\begin{figure}[!b]
	\begin{center}
	\includegraphics[scale=0.77]{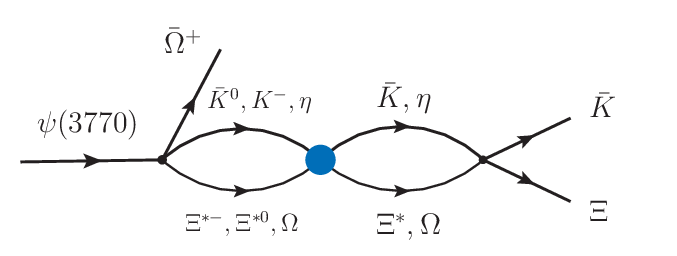}
	\end{center}
	\vspace{-0.7cm}
	\caption{Reaction mechanism for the $\psi(3770)\to \bar{\Omega}^+ \bar{K} \Xi$ decay. {The blue thick dot indicates the resonant $t_{ij}$ scattering matrix. The $\bar K \Xi^* \to \bar K \Xi$ and $\eta \Omega \to \bar K \Xi$ vertices correspond to the transition potential between these channels \cite{Pavao:2018xub}.}}
	\label{feynDiag1}
\end{figure}

Once this is done, we must allow the $\bar{K}\Xi^*, \eta\Omega$ states to interact and finally undergo a transition to $\bar{K}\Xi$. 
This is depicted diagrammatically in Fig.~\ref{feynDiag1}. 
This mechanism requires the use of the Clebsch-Gordan coefficients, the $\bar{K}\Xi^*$ and $\eta\Omega$ propagation and the transition $t$-matrices between these channels, plus the couplings of $\bar{K}\Xi^*$ and $\eta\Omega$ to $\bar{K}\Xi$, including explicitly a $D$-wave transition, which are given explicitly in Ref.~\cite{Pavao:2018xub} (see Table~I of Ref.~\cite{Pavao:2018xub}) as $\alpha\,q_\text{on}^2, \;\beta\,q_\text{on}^2$, respectively, with $q_\text{on}$ given by
\begin{align}
    q_\text{on}=\frac{\lambda^{1 / 2}\left(s, m_K^2, M_{\Xi}^2\right)}{2 \sqrt{s}},
\end{align}
with $\alpha=4.0\times10^{-8} \mev$, $\beta=1.5\times10^{-8} \mev$ from Table~I of Ref.~\cite{Pavao:2018xub}.

We also work in isospin basis of the states to connect with the work of Ref.~\cite{Pavao:2018xub} where one works with the isospin basis. 
Since
\begin{align}
\vert\bar{K} \Xi^{*}, I=0 \rangle =\frac{1}{\sqrt{2}} \vert \bar{K}^0 \Xi^{*-}+ {K}^- \Xi^{*0}\rangle, 
\label{eq:iso}
\end{align}
both $\bar{K}^0 \Xi^{*-}$ and ${K}^- \Xi^{*0}$ have a component $\frac{1}{\sqrt{2}}$ of the $\vert\bar{K} \Xi^{*}, I=0 \rangle$ state. 
Then we have for the transition amplitude, summing the contributions of the internal states, $\bar{K}^0 \Xi^{*-},\;{K}^- \Xi^{*0}$ and $\eta\Omega$,
\begin{align}\label{tmatirxtotal}
 t=\nonumber
 &\Bigg[\dfrac{1}{2} \,G_{\bar{K} \Xi^*}\; \dfrac{1}{\sqrt{2}}\, t_{\bar{K} \Xi^*, \bar{K} \Xi^*} \,G_{\bar{K} \Xi^*}\;\alpha\,q_\text{on}^2  \nonumber \\
 &+ 
 \dfrac{1}{2} \, G_{\bar{K} \Xi^*} \,\dfrac{1}{\sqrt{2}} \, t_{\bar{K} \Xi^*, \bar{K} \Xi^*} \, G_{\bar{K} \Xi^*}\;\alpha\,q_\text{on}^2 \nonumber \\
 & + \dfrac{1}{\sqrt{2}} \, G_{\eta_\Omega} \, t_{\eta_\Omega, \bar{K} \Xi^*} \, G_{\bar{K} \Xi^*} \;\alpha\,q_\text{on}^2 \nonumber \\
&+ \dfrac{1}{2} \, G_{\bar{K} \Xi^*} \,\dfrac{1}{\sqrt{2}} \, t_{\bar{K} \Xi^*, \eta_\Omega} \, G_{\eta_\Omega}\;\beta \,q_\text{on}^2 \nonumber \\
&+ 
\dfrac{1}{2} \,G_{\bar{K} \Xi^*} \,\dfrac{1}{\sqrt{2}} \, t_{\bar{K} \Xi^*, \eta_\Omega} \, G_{\eta_\Omega} \;\beta\,q_\text{on}^2 \nonumber\\
&+
\dfrac{1}{\sqrt{2}}\, G_{ \eta_\Omega} \,t_{\eta_\Omega,\eta_\Omega} \,G_{\eta_\Omega} \; \beta \,q_\text{on}^2\Bigg]\;\vec{\epsilon}\,(\psi)\cdot \vec p_{\bar{\Omega}} \nonumber\\ 
=&~t'\;\vec{\epsilon}\,(\psi)\cdot \vec p_{\bar{\Omega}},
\end{align}
where $t'$ corresponds to the expression in brackets. 
The term  $\vec{\epsilon}\,(\psi)\cdot \vec p_{\bar{\Omega}}$ is included to ensure parity conservation.
The first coefficient in the terms of Eq.~\eqref{tmatirxtotal} corresponds to the weights of the Clebsch-Gordan coefficients of Table~\ref{tableI}, and the second to the $I=0$ isospin content of the $\bar K^0 \Xi^{*-}$ and $K^-\Xi^{*0}$ intermediate states of the first loop.
In the second loop, the states are in isospin basis.

With this amplitude we obtain the $\bar{K}\Xi$ mass distribution for the $\psi(3770)\to \bar{\Omega}^+ \bar{K}^- \Xi^{0}$ decay, when summing and averaging over polarizations, as,
\begin{align}\label{massdis1}
	\nonumber
	\dfrac{\dd \Gamma}{\dd M_\text{inv}(\bar{K}\Xi)}&=\dfrac{1}{(2\pi)^3} \; \dfrac{1}{4M_\psi^2}\; p_{\bar{\Omega}}\, \tilde{k}\;\bar{\sum}\sum|t|^2 \\
	&=\dfrac{1}{(2\pi)^3}\; \dfrac{1}{4\,M_\psi^2}\; p_{\bar{\Omega}}\, \tilde{k}\; |t'|^2 \, \dfrac{1}{3}\, p_{\bar{\Omega}}^{~2},
\end{align}
with the momenta $p_{\bar{\Omega}}$ and 
$\tilde{k}$ given by
\begin{equation}
\begin{split}
p_{\bar{\Omega}}=\dfrac{\lambda^{1 / 2}\left(M_\psi^2, M_{\bar{\Omega}}^2, M_\text{inv}^2\left(\bar{K} \Xi\right)\right)}{2 M_\psi}, \\
\tilde{k}=\dfrac{\lambda^{1 / 2}\left(M_{\operatorname{inv}}^2\left(\bar{K} \Xi\right), m_{\bar{K}}^2, {M}_\Xi^2\right)}{2 M_{\operatorname{inv}}\left(\bar{K} {\Xi}\right)}.
\end{split}
\end{equation}
Note that by working in the isospin basis, we are automatically summing over the different charges of $\bar K \Xi$ in $I=0$.
If one wishes to know the fractions into $\bar K^0 \Xi^-$ and $K^- \Xi^0$, one simply uses the Clebsch-Gordan coefficients (the same as in Eq.~\eqref{eq:iso}) and finds that there would be equal strength in the two charged channels.
Furthermore, the loop functions in Eq.~\eqref{tmatirxtotal} are given by
\begin{align}\label{eq:Gi}
 G_i(\sqrt{s}) 
 =&\int_{|\vec{q}\,|<q_\mathrm{max}} \dfrac{\dd^3 q}{(2 \pi)^3} \; \dfrac{1}{2 \omega_i(\vec{q}\,)} \; \dfrac{M_i}{E_i(\vec{q}\,)} \nonumber \\
 &~~~~~~~~~\times \frac{1}{\sqrt{s}-\omega_i(\vec{q}\,)-E_i(\vec{q}\,)+i \epsilon},
\end{align}
where $M_i$ ($m_i$) stands for the mass of the baryon (meson), and $w_i(\vec q\,)=\sqrt{m_i^2+\vec{q}^{\;2}}$, $E_i(\vec q\,)=\sqrt{M_i^2+\vec{q}^{\;2}}$. 
In addition, $q_{\max }= 735 \mev$ is the cutoff used in Ref.~\cite{Pavao:2018xub}.
We should note that the $t_{ij}$ matrices used here actually have a momentum dependence \cite{Gamermann:2009uq}
\begin{equation*}
	t_{ij}(\vec q, \vec q\,'\,)=\Theta (q_{\rm max}-|\vec q\;|)\; \Theta (q_{\rm max}-|\vec q\,'|),
\end{equation*}
which is in consistency with the cutoff regularization used in the $G$ function.
Then the two loops in Fig.~\ref{feynDiag1} result with the $G$ function regularized as in Eq.~\eqref{eq:Gi}.
Note the need for the two loops if we want to generate the resonance from the $\bar K \Xi^*$ and $\eta \Omega$ channels, and then make a transition from this resonance to the $\bar K \Xi$ channel which proceeds in $D$-wave.
The loop to the left in Fig.~\ref{feynDiag1} connects to $\bar K \Xi^*, \eta \Omega$ channels through the resonant scattering matrix.
The second loop connects the $\bar K \Xi^*, \eta \Omega$ to $\bar K \Xi$ through the transition potential of Ref.~\cite{Pavao:2018xub}. 
Since in Ref.~\cite{Pavao:2018xub} the couplings of the $\Omega(2012)$ resonance to the $\bar{K}\Xi^*$ and $\eta\Omega$ states are given and we are working close to the $\Omega(2012)$ pole,
we take advantage to write the transition amplitudes as
\begin{align}
& t_{i j}=\dfrac{g_i \, g_j}{\sqrt{s}-M_{\Omega(2012)}+\dfrac{i \Gamma_{\Omega(2012)}}{2}} .
\end{align}
The arguments of $G$ and $t_{i j}$ are $\sqrt{s}\equiv M_{\operatorname{inv}}\left(\bar{K} {\Xi}\right)$.  
For the different channels we have
\begin{align}\label{tmatrix_diff_chan}
& t_{\bar{K} \Xi^*, \bar{K} \Xi^*}=\dfrac{g_{\bar{K} \Xi^*} ~ g_{ \bar{K} \Xi^*}}{M_{\operatorname{inv}}\left(\bar{K} {\Xi}\right)-M_{\Omega(2012)}+\dfrac{i \Gamma_{\Omega(2012)}}{2}},\nonumber \\
& t_{\bar{K} \Xi^*, \eta_\Omega}=\dfrac{g_{\bar{K} \Xi^*} ~g_{\eta_\Omega}}{M_{\operatorname{inv}}\left(\bar{K} {\Xi}\right)-M_{\Omega(2012)}+\dfrac{i \Gamma_{\Omega(2012)}}{2}},\\
& t_{\eta_\Omega,\eta_\Omega}=\dfrac{g_{\eta_\Omega} ~g_{ \eta_\Omega}}{M_{\operatorname{inv}}\left(\bar{K} {\Xi}\right)-M_{\Omega(2012)}+\dfrac{i \Gamma_{\Omega(2012)}}{2}}, \nonumber
\end{align}
with $g_i$, the couplings of $\Omega(2012)$  to the  different channels listed in Table~II of Ref.~\cite{Pavao:2018xub}, whose values are given by $g_{\bar{K} \Xi^*}=2.01 +i\; 0.02$, $g_{\eta_\Omega}=2.84 - i\;0.01$, $g_{ \bar{K} \Xi}=-0.29+i \; 0.04$. 
According to the PDG~\cite{ParticleDataGroup:2024cfk}, the $\Omega(2012)$ mass is $2012.4 \mev$, and its width is $\Gamma_{\Omega(2012)} = 6.4 \mev$.

\begin{figure}[t]
	\begin{center}
	\includegraphics[scale=0.75]{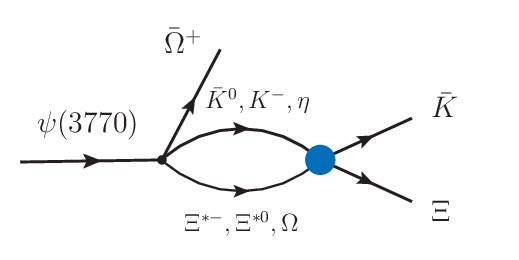}
	\end{center}
	\vspace{-0.7cm}
	\caption{The simplified mechanism of the reaction $\psi(3770)\to \bar{\Omega}^+ \bar{K} \Xi$.
	The blue thick dot indicates the resonant transition matrix from $\bar K \Xi^*\; (\eta \Omega)$ to $\bar K \Xi$ \cite{Pavao:2018xub}.}
	\label{feynDiag2}
\end{figure}

There is an alternative method to evaluate Eq.~\eqref{tmatirxtotal}.
Since the effective couplings of $\Omega(2012)$ to $\bar{K}\Xi$, including the $D$-wave  factor, is also calculated in Ref.~\cite{Pavao:2018xub}, we can rewrite Eq.~(\ref{tmatirxtotal}), as shown in Fig.~\ref{feynDiag2}, in a simpler form as follows,
\begin{align}\label{tsimle}
& \tilde{t}'=\dfrac{1}{\sqrt{2}} \;G_{\bar{K} \Xi^*} \; t_{\bar{K} \Xi^*, \bar{K} \Xi}  +\dfrac{1}{\sqrt{2}} \;G_{ \eta_\Omega} \;t_{\eta_\Omega,\bar{K} \Xi}, 
\end{align}
with $t_{ij}$ given by
\begin{align}
& t_{\bar{K} \Xi^*,\bar{K} \Xi}=\dfrac{g_{\bar{K} \Xi^*} ~g_{ \bar{K} \Xi}}{M_{\operatorname{inv}}\left(\bar{K} {\Xi}\right)-M_{\Omega(2012)}+\dfrac{i \Gamma_{\Omega(2012)}}{2}},\\
& t_{\eta_\Omega,\bar{K} \Xi}=\dfrac{g_{\eta_\Omega} ~g_{ \bar{K} \Xi}}{M_{\operatorname{inv}}\left(\bar{K} {\Xi}\right)-M_{\Omega(2012)}+\dfrac{i \Gamma_{\Omega(2012)}}{2}}.
\end{align}

\subsection{$\psi(3770)\to \bar{\Omega}^+\bar{K}\Xi^* \to \bar{\Omega}^+\bar{K} \pi\Xi$}

\begin{figure}[b]
  \centering
    \includegraphics[scale=0.75]{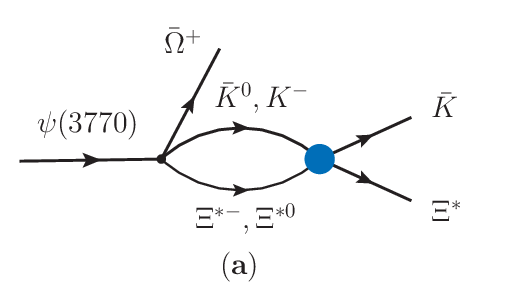}
	\includegraphics[scale=0.75]{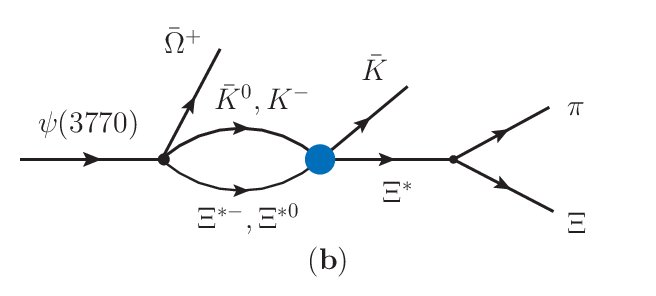}
  \caption{(a) Mechanism of the reaction $\psi(3770)\to \bar{\Omega}^+ \bar{K}\Xi^*$, and (b) the one for the $\psi(3770)\to \bar{\Omega}^+ \bar{K} \pi\Xi$.}
   \label{feynDiag3}
\end{figure}

The reaction $\psi(3770)\to \bar{\Omega}^+\bar{K}\Xi^*$ proceeds now according to the diagram in Fig.~\ref{feynDiag3}(a). 
The amplitude corresponding to the process in Fig.~\ref{feynDiag3}(a) is given by
 \begin{align}
& t^{\prime}=\dfrac{1}{\sqrt{2}} \;G_{\bar{K} \Xi^*} \;t_{\bar{K} \Xi^*, \bar{K} \Xi^*}  +\dfrac{1}{\sqrt{2}}\; G_{ \eta_\Omega}\; t_{\eta_\Omega,\bar{K} \Xi^*}.
\end{align}
In Fig.~\ref{feynDiag3}(b) we allow the $\Xi^*$ to decay to $\pi\Xi$ and we have then  a $4$-body decay mode. 
Once again, $t_{ij}$ are given by
 \begin{align}
& t_{\bar{K} \Xi^*, \bar{K} \Xi^*}=\dfrac{g_{\bar{K} \Xi^*} ~g_{\bar{K} \Xi^*}}{M_{\operatorname{inv}}\left(\bar{K} {\Xi}^*\right)-M_{\Omega(2012)}+\dfrac{i \Gamma_{\Omega(2012)}}{2}},\\
& t_{\eta_\Omega,\bar{K} \Xi^*}=\dfrac{g_{\eta_\Omega} ~g_{\bar{K} \Xi^*}}{M_{\operatorname{inv}}\left(\bar{K} {\Xi}^*\right)-M_{\Omega(2012)}+\dfrac{i \Gamma_{\Omega(2012)}}{2}}.
\end{align}
The $\bar{K} \Xi^*$ mass distribution for the decay $\psi(3770)\to \bar{\Omega}^+ \bar{K}\Xi^* $ is then given by
\begin{align}
    \dfrac{\dd \Gamma}{\dd M_\text{inv}(\bar{K}\Xi^*)}=\dfrac{1}{(2\pi)^3}\;\dfrac{1}{4M_\psi^2}\; p_{\bar{\Omega}}\; \tilde{k}^\prime\; |t'|^2\; \dfrac{1}{3}\, {p}_{\bar{\Omega}}^{~2},
\end{align}
where $\tilde{k}'$ is the momentum of the $\bar{K}$ in the $\bar{K} \Xi^*$ rest frame. 
Furthermore,
$p_{\bar{\Omega}}$ and $\tilde{k}^\prime$ are given by
\begin{equation}
\begin{split}
p_{\bar{\Omega}}=\dfrac{\lambda^{1 / 2}\left(M_\psi^2, M_{\bar{\Omega}}^2, M_\text{inv}^2\left(\bar{K} \Xi^*\right)\right)}{2 M_\psi}, \\
\tilde{k}^\prime=\dfrac{\lambda^{1 / 2}\left(M_{\operatorname{inv}}^2\left(\bar{K} \Xi^*\right), m_{\bar{K}}^2, {M}_{\Xi^*}^2\right)}{2 M_{\operatorname{inv}}\left(\bar{K} {\Xi}^*\right)}.
\end{split}
\end{equation}

We can go forward to consider the mass distribution of the $\Xi^*$ from the decay $\Xi^* \to \pi \Xi$. This is shown in Fig.~\ref{feynDiag3}(b). 
Considering the  $\Xi^*$ mass distribution from the $\pi\Xi$ decay ($\Xi^*$ spectral function), and that the branching ratio of $\Xi^*\to \pi\Xi$ is $100\%$. 
We can write
\begin{align}\label{ttotal_2}
    &\dfrac{\dd \Gamma}{\dd M_\text{inv}(\bar{K}\Xi^*) \;\dd\tilde{M}_\text{inv}} \nonumber \\
	&=\dfrac{1}{(2\pi)^3}\, \dfrac{1}{4M_\psi^2}\; p_{\bar{\Omega}} \, \tilde{k}^{\prime}\; |t^{\prime}|^2 \;\dfrac{1}{3} \, p_{\bar{\Omega}}^{~2}\nonumber \\
	&~\times (-\dfrac{1}{\pi}) \, \Im\dfrac{1}{\tilde{M}_\text{inv}-M_{\Xi^*}+i\;\Gamma_{\Xi^*}\left(\tilde{M}_{\text {inv}}\right)/2},
\end{align}
where now
\begin{align}
& \tilde{k}^\prime=\dfrac{\lambda^{1 / 2}\left(M_{\operatorname{inv}}^2\left(\bar{K} \Xi^*\right), m_{\bar{K}}^2, \tilde{M}_\text {inv }^2\right)}{2 M_{\operatorname{inv}}\left(\bar{K} {\Xi}^*\right)}.
\end{align}
In addition, we take the $\Xi^*$ width as energy dependent,
\begin{align}
&\Gamma_{\Xi^*}\left(\tilde{M}_{\text {inv }}\right)={\Gamma}_{\Xi^*,\rm on}\;\left(\dfrac{p_\pi}{p_{\pi, \text {on}}}\right)^3  \dfrac{M_{\Xi^*}}{\tilde{M}_{\text {inv}}}, \\
& p_\pi=\dfrac{\lambda^{V_2}\left(\tilde{M}_{\text {inv }}^2, m_{\pi}^2, {M}_\Xi^2\right)}{2 \tilde{M}_\text {inv }}, \\
& p_{\pi, \text {on}}=\dfrac{\lambda^{1 / 2}\left(M_{\Xi^*}^2, m_\pi^2, M_{\Xi}^2\right)}{2 M_{\Xi^*}},
\end{align}
where ${\Gamma}_{\Xi^*,\rm on}$ corresponds to the $\Xi^{*}$ width from the PDG, ${\Gamma}_{\Xi^*,\rm on}=9.1 \mev$ \cite{ParticleDataGroup:2024cfk}, 
while ${p}_{\pi}$ is the momentum of the $\pi$ in the decay of a $\Xi^{*}$, with invariant mass $M_{\mathrm{inv}}(\Xi^{*})=\tilde{M}_{\mathrm{inv}}$, into $\pi \Xi$. 
Additionally, ${p}_{\pi,\text{on}}$ stands for the same momentum for the nominal $\Xi^{*}$ mass.

Once again, by means of Eq.~\eqref{ttotal_2} one is considering all the possible charged states of $\bar K^0 \Xi^{*-} \,(\pi^- \Xi^0, \, \pi^0 \Xi^-)$ and $K^- \Xi^{*0} \,(\pi^0 \Xi^0,\, \pi^+ \Xi^-)$.
Using Clebsch-Gordan coefficients, it is trivial to see the relative weights of each charge states:
$\dfrac{1}{6}$ for $\bar K^0 \pi^0 \Xi^-$, $\dfrac{1}{3}$ for $\bar K^0 \pi^- \Xi^0$, $\dfrac{1}{3}$ for $K^- \pi^+ \Xi^-$ and $\dfrac{1}{6}$ for $K^- \pi^0 \Xi^0$.

\section{results}

\begin{figure}[!b]
\centering
\includegraphics[width=0.51\textwidth]{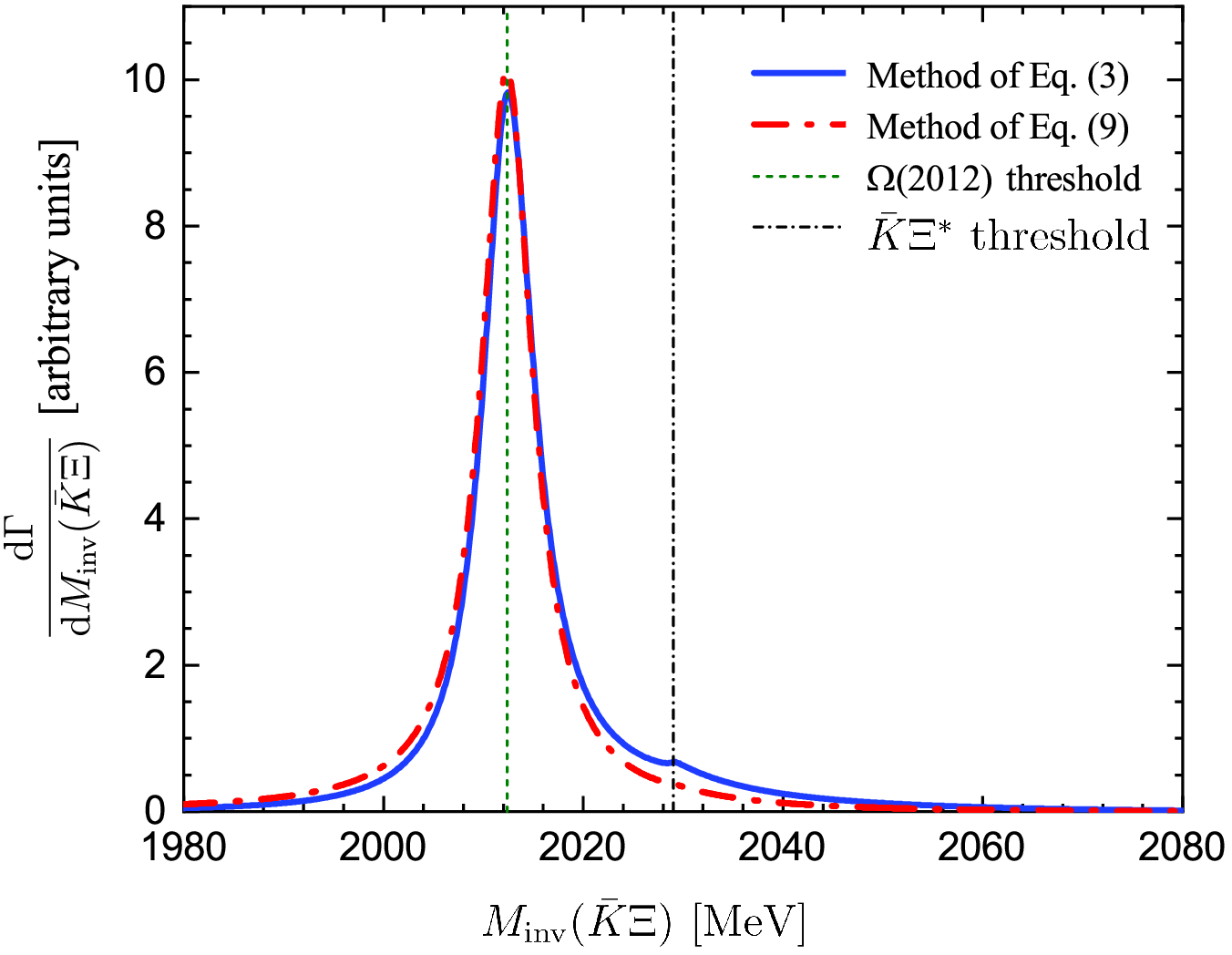}
   \caption{The mass distribution  $\frac{\dd\Gamma}{\dd M_{\mathrm{inv}}(\bar{K}\Xi)}$ of $\psi(3770)\to \bar{\Omega}\bar{K}\Xi$ as a function of $M_{\mathrm{inv}}(\bar{K}\Xi)$. The $\frac{\dd\Gamma}{\dd M_{\mathrm{inv}}(\bar{K}\Xi)}$ values obtained with two different methods are shown by different lines (see text). The vertical green dashed line and black short-dashed-dotted line represent the threshold of $\Omega(2012)$ and the boundary of $\bar{K}\Xi^*$, respectively.}
    \label{res1}
\end{figure}

\begin{figure}[!t]
\centering
\includegraphics[width=0.51\textwidth]{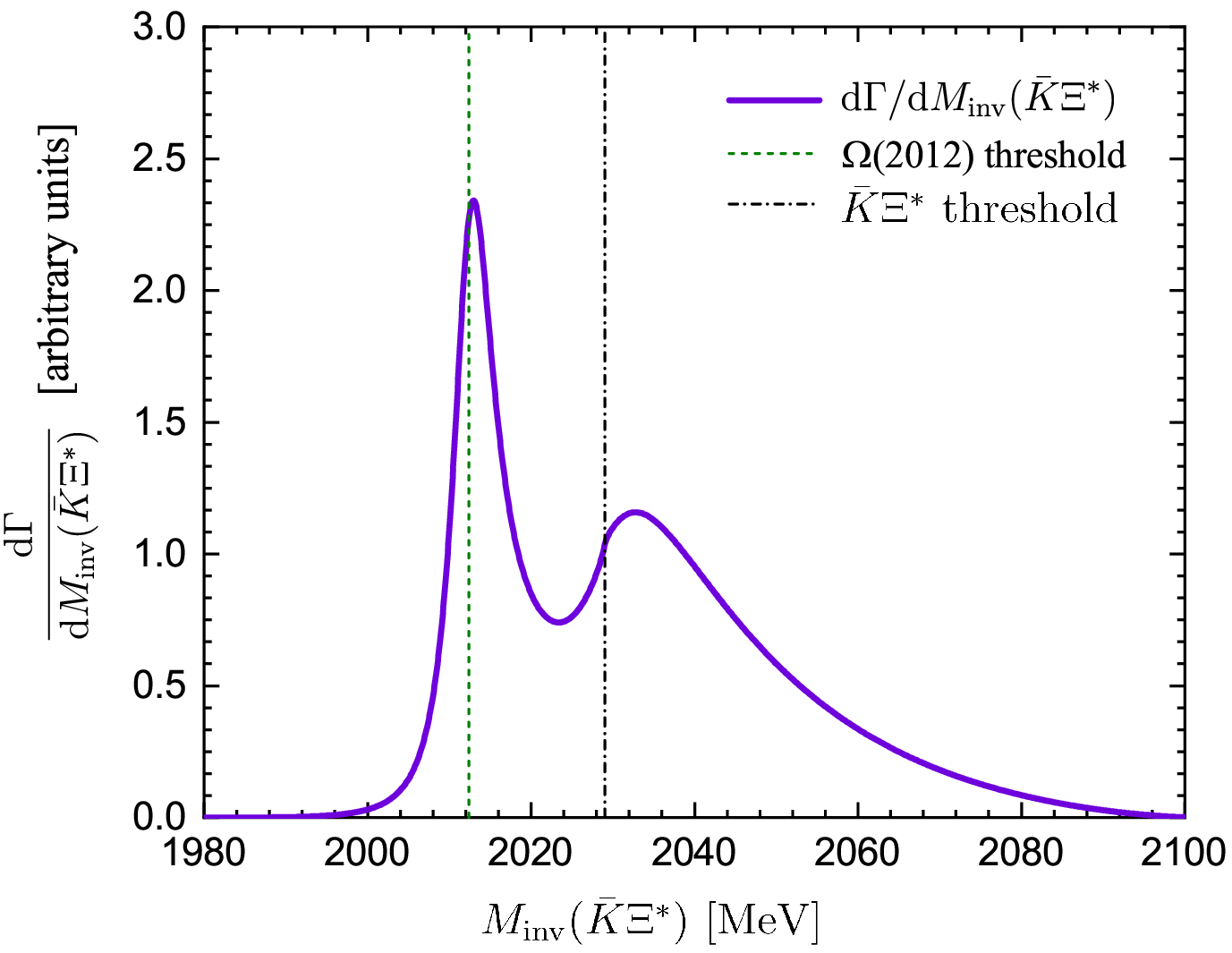}
\caption{The mass distribution of $\frac{\dd \Gamma}{\dd M_{\mathrm {inv}}(\bar{K}\Xi^*)}$ as a function of $M_{\mathrm {inv}}(\bar{K}\Xi^*)$, for the $\psi(3770)\to \bar{\Omega}\bar{K}\Xi^* \;(\pi\Xi)$, obtained by integrating over $\tilde{M}_{\mathrm {inv}}(\Xi^*)$ in Eq.~\eqref{ttotal_2}. 
The labels are the same as in Fig.~\ref{res1}.}
\label{res2}
\end{figure}

We show the results for the $\psi(3770) \to \bar{\Omega} \bar{K} \Xi$ reaction in Fig.~\ref{res1}, and for the $\psi(3770) \to \bar{\Omega} \bar{K} \Xi^* (1530) (\bar{\Omega} \bar{K} \pi \Xi)$ reaction in Fig.~\ref{res2}.
In Fig.~\ref{res1}, we plot $\dd \Gamma/\dd M_{\mathrm{inv}}(\bar{K}\Xi)$ for the $\psi(3770)$ decay to $\bar{\Omega}\bar{K} \Xi$. 
We see that there is a clear peak for the $\Omega(2012)$ excitation, with the right position and width of the resonance.  
We also show the results obtained with the two methods proposed in Eqs.~\eqref{tmatirxtotal} and \eqref{tsimle}. 
As we can see, the results of both methods are remarkably similar.  
We do not give the absolute value of the strength of the signal, because there is an arbitrary normalization constant multiplying the amplitudes, which we ignore.
Yet, since the mechanism used here and the couplings of the $\psi(3770)$ to the $\bar{\Omega} \bar{K} \Xi^*(1530)$ through the SU(3) Clebsch-Gordan coefficients are similar to those entering the evaluation of the $\psi(3686)$ to $\bar{\Xi} \bar{K} \Sigma^*(1385)$ producing the $\Xi(1820)$ resonances in Ref.~\cite{Duan:2024ygq}, 
we expect that the absolute rates are similar to those seen in the BESIII experiment~\cite{BESIII:2023mlv}.

In Fig.~\ref{res2}, we show the results for the reaction $\psi(3770) \to \bar{\Omega} \bar{K} \Xi^*(1530) \to \bar{\Omega} \bar{K} \Xi \pi$. 
We show results for $\dd \Gamma/ \dd M_{\mathrm{inv}}(\bar{K} \Xi^*)$, which we obtain by integrating Eq.~\eqref{ttotal_2} over $\tilde{M}_{\mathrm{inv}}$. 
The results are a bit surprising. 
We obtain two peaks, a narrow peak for the $\Omega(2012)$ excitation, and a second, broader peak, that requires some attention. 
This peak originates from the opening of the $\bar{K} \Xi^*$ state to $\Omega(2012)$ decay, occurring just around and slightly above the $\bar{K} \Xi^*$ threshold.
The width of the peak reflects the width of the $\Xi^*(1530)$ state. 
This peak is then tied to the structure of the $\Omega(2012)$ as a molecular state formed from $\bar{K} \Xi^*$ and $\eta \Omega$.  
This is interesting in itself and it is not the first time that such an effect is reported. 
Indeed, in Ref.~\cite{Liang:2020jtw} it was found that the ``$f_1(1420)$" state came naturally as a consequence of the decay of the $f_1(1285) $ to the $K \bar{K}^*$, with the $f_1(1285) $ state being a molecular state of the $K \bar{K}^*$ and $\bar{K} K^*$ components. 
In this sense it was claimed in Ref.~\cite{Liang:2020jtw} that the $f_1(1420)$ was not a real resonance but a manifestation of the decay of the $f_1(1285)$ into the $K \bar{K}^*$ channel (see also related work in Refs.~\cite{Debastiani:2016xgg,Lin:2023ajb}). 
In order to test the molecular nature of the $\Omega (2012)$ it would be very interesting to observe this second peak obtained here. 
This, in addition would shed further light on the origin of the $f_1(1420) $ resonance.
Actually, it is very interesting to call the attention here to the experimental work of the Belle experiment of Ref.~\cite{Belle:2022mrg}. 
Indeed, in Fig.~3(a) of that latter work there seems to be a small peak at around $2025 \mev$, close to the energy $2033 \mev$ where we find the second peak in Fig.~\ref{res2}. \footnote{Figure~\ref{res2} is calculated using average masses of $\Xi^*$ and $\bar K$. If we use the physical masses of $\Xi^{*0}$ and $K^-$ to adapt to the situation of Ref.~\cite{Belle:2022mrg}, the second peak moves about $4 \mev$ to lower energies, in better agreement with the results of Ref.~\cite{Belle:2022mrg}.}
The statistics is not sufficient to make conclusions, but the finding of the present work should provide sufficient motivation to look at that in detail with more statistics. 
Similarly, the observation of the second peak on the $\psi(3770)\to \bar{\Omega}\bar{K}\pi\Xi$ reaction would be very valuable in the search to unravel the nature of the $\Omega(2012)$.

Coming back to Fig.~\ref{res2}, it would be most interesting to perform the proposed experiment with sufficient statistics to eventually observe the two peaks predicted here, 
based on the molecular nature of the $\Omega(2012)$. 
This would give boost to the molecular picture of this resonance. 
Concerning the absolute rate of the reaction, we have here the same normalization constant than in the former reaction, 
which means that the relative values of the mass distributions in Figs.~\ref{res1} and \ref{res2} are a genuine prediction of the theory, 
and something that can also be tested experimentally.

\section{conclusion}
We have theoretically investigated on the $\psi(3770)$ decays to the $\bar{\Omega}\bar{K} \Xi$ and 
$\bar{\Omega}\bar{K} \Xi^*(1530)$ ($\bar{\Omega}\bar{K} \pi \Xi$) channels. 
Based on the assumption that the  $\Omega(2012)$ is a molecular state of the $\bar{K} \Xi^*(1530)$ and $\eta \Omega$, 
as supported by a Belle experiment and  many theoretical works, 
we have evaluated the mass distributions for these two reactions. 
Our results show clear peaks corresponding to the  $\Omega(2012)$ resonance. 
Although the absolute rates are not computed, we rely upon the similar reaction studied at BESIII for the $\psi(3686) \to \bar{\Xi} \bar{K} \Lambda$, which indicates the feasibility of the proposed reaction in the same facility, 
and in future updates, with hopefully very good statistics to observe the structures obtained here. 

A remarkable finding of the present work is the existence of two peaks in the $\psi(3770)$ decay to $\bar{\Omega} \bar{K} \Xi^*(1530)$ ($\bar{\Omega}\bar{K} \pi \Xi$).  
There is a narrow peak corresponding to the $\Omega(2012)$ excitation and a broader peak at $2035 \mev$, 
that comes from the opening of the $\bar{K} \Xi^*(1530)$ channel for the $\Omega(2012)$ decay, 
something that only happens because of the large overlap of the $\Omega(2012)$ to the $\bar{K} \Xi^*(1530)$ molecular component. 
The observation of this second peak would provide a significant support to the molecular picture of the $\Omega(2012)$.
The findings of the present work should provide a clear incentive to perform the experimental reactions, 
something feasible at BESIII, particularly in further updates that would allow to have very large statistics. 

\section{Acknowledgments}
We thank Prof. Cheng-Ping Shen for useful discussions. 
This work is partly supported by the National Natural Science Foundation of China under Grants  No. 12405089 and No. 12247108 and the China Postdoctoral Science Foundation under Grant No. 2022M720360 and No. 2022M720359. 
This work is also supported by
the Spanish Ministerio de Economia y Competitivi-dad (MINECO) and European FEDER funds under Contracts No. FIS2017-84038-C2-1-P B, PID2020-112777GB-I00, 
and by Generalitat Valenciana under contract PROMETEO/2020/023. 
This project has received funding from the European Union Horizon 2020 research and innovation programme under the program H2020-INFRAIA-2018-1, grant agreement No. 824093 of the STRONG-2020 project. 
This work is supported by the Spanish Ministerio de Ciencia e Innovaci\'on (MICINN) under contracts PID2020-112777GB-I00, PID2023-147458NB-C21 and CEX2023-001292-S; by Generalitat Valenciana under contracts PROMETEO/2020/023 and  CIPROM/2023/59. 
This work is partly supported by the National Natural Science Foundation of China (NSFC) under Grants No. 12365019 and No. 11975083, 
and by the Central Government Guidance Funds for Local Scientific and Technological Development, China (No. Guike ZY22096024), 
the Natural Science Foundation of Guangxi province under Grant No. 2023JJA110076, 
and partly by the Natural Science Foundation of Changsha under Grant No. kq2208257 and the Natural Science Foundation of Hunan province under Grant No. 2023JJ30647 (CWX).
J. M. Dias acknowledges the support from the Chinese Academy of Sciences under Grants No. XDB34030000 and No. YSBR-101; by the National Key R\& D Program of China under Grant No. 2023YFA1606703.

\bibliographystyle{a}
\bibliography{ref}

\begin{thebibliography}{32}%
\makeatletter
\providecommand \@ifxundefined [1]{%
 \@ifx{#1\undefined}
}%
\providecommand \@ifnum [1]{%
 \ifnum #1\expandafter \@firstoftwo
 \else \expandafter \@secondoftwo
 \fi
}%
\providecommand \@ifx [1]{%
 \ifx #1\expandafter \@firstoftwo
 \else \expandafter \@secondoftwo
 \fi
}%
\providecommand \natexlab [1]{#1}%
\providecommand \enquote  [1]{``#1''}%
\providecommand \bibnamefont  [1]{#1}%
\providecommand \bibfnamefont [1]{#1}%
\providecommand \citenamefont [1]{#1}%
\providecommand \href@noop [0]{\@secondoftwo}%
\providecommand \href [0]{\begingroup \@sanitize@url \@href}%
\providecommand \@href[1]{\@@startlink{#1}\@@href}%
\providecommand \@@href[1]{\endgroup#1\@@endlink}%
\providecommand \@sanitize@url [0]{\catcode `\\12\catcode `\$12\catcode `\&12\catcode `\#12\catcode `\^12\catcode `\_12\catcode `\%12\relax}%
\providecommand \@@startlink[1]{}%
\providecommand \@@endlink[0]{}%
\providecommand \url  [0]{\begingroup\@sanitize@url \@url }%
\providecommand \@url [1]{\endgroup\@href {#1}{\urlprefix }}%
\providecommand \urlprefix  [0]{URL }%
\providecommand \Eprint [0]{\href }%
\providecommand \doibase [0]{https://doi.org/}%
\providecommand \selectlanguage [0]{\@gobble}%
\providecommand \bibinfo  [0]{\@secondoftwo}%
\providecommand \bibfield  [0]{\@secondoftwo}%
\providecommand \translation [1]{[#1]}%
\providecommand \BibitemOpen [0]{}%
\providecommand \bibitemStop [0]{}%
\providecommand \bibitemNoStop [0]{.\EOS\space}%
\providecommand \EOS [0]{\spacefactor3000\relax}%
\providecommand \BibitemShut  [1]{\csname bibitem#1\endcsname}%
\let\auto@bib@innerbib\@empty
\bibitem [{\citenamefont {Yelton}\ {\it et~al.}(2018)\citenamefont {Yelton} {\it et~al.}}]{Belle:2018mqs}%
  \BibitemOpen
  \bibfield  {author} {\bibinfo {author} {\bibfnamefont {J.}~\bibnamefont {Yelton}} {\it et~al.} (\bibinfo {collaboration} {Belle}),\ }\bibinfo {title} {{Observation of an Excited $\Omega^-$ Baryon}},\ \href {https://doi.org/10.1103/PhysRevLett.121.052003} {\bibfield  {journal} {\bibinfo  {journal} {Phys. Rev. Lett.}\ }\textbf {\bibinfo {volume} {121}},\ \bibinfo {pages} {052003} (\bibinfo {year} {2018})},\ \Eprint {https://arxiv.org/abs/1805.09384} {arXiv:1805.09384 [hep-ex]} \BibitemShut {NoStop}%
\bibitem [{\citenamefont {Hofmann}\ and\ \citenamefont {Lutz}(2006)}]{Hofmann:2006qx}%
  \BibitemOpen
  \bibfield  {author} {\bibinfo {author} {\bibfnamefont {J.}~\bibnamefont {Hofmann}}\ and\ \bibinfo {author} {\bibfnamefont {M.~F.~M.}\ \bibnamefont {Lutz}},\ }\bibinfo {title} {{D-wave baryon resonances with charm from coupled-channel dynamics}},\ \href {https://doi.org/10.1016/j.nuclphysa.2006.07.004} {\bibfield  {journal} {\bibinfo  {journal} {Nucl. Phys. A}\ }\textbf {\bibinfo {volume} {776}},\ \bibinfo {pages} {17} (\bibinfo {year} {2006})},\ \Eprint {https://arxiv.org/abs/hep-ph/0601249} {arXiv:hep-ph/0601249} \BibitemShut {NoStop}%
\bibitem [{\citenamefont {Sarkar}\ {\it et~al.}(2005)\citenamefont {Sarkar}, \citenamefont {Oset},\ and\ \citenamefont {Vicente~Vacas}}]{Sarkar:2004jh}%
  \BibitemOpen
  \bibfield  {author} {\bibinfo {author} {\bibfnamefont {S.}~\bibnamefont {Sarkar}}, \bibinfo {author} {\bibfnamefont {E.}~\bibnamefont {Oset}},\ and\ \bibinfo {author} {\bibfnamefont {M.~J.}\ \bibnamefont {Vicente~Vacas}},\ }\bibinfo {title} {{Baryonic resonances from baryon decuplet-meson octet interaction}},\ \href {https://doi.org/10.1016/j.nuclphysa.2005.01.006} {\bibfield  {journal} {\bibinfo  {journal} {Nucl. Phys. A}\ }\textbf {\bibinfo {volume} {750}},\ \bibinfo {pages} {294} (\bibinfo {year} {2005})},\ \bibinfo {note} {[Erratum: Nucl.Phys.A 780, 90--90 (2006)]},\ \Eprint {https://arxiv.org/abs/nucl-th/0407025} {arXiv:nucl-th/0407025} \BibitemShut {NoStop}%
\bibitem [{\citenamefont {Valderrama}(2018)}]{Valderrama:2018bmv}%
  \BibitemOpen
  \bibfield  {author} {\bibinfo {author} {\bibfnamefont {M.~P.}\ \bibnamefont {Valderrama}},\ }\bibinfo {title} {{$\Omega(2012)$ as a hadronic molecule}},\ \href {https://doi.org/10.1103/PhysRevD.98.054009} {\bibfield  {journal} {\bibinfo  {journal} {Phys. Rev. D}\ }\textbf {\bibinfo {volume} {98}},\ \bibinfo {pages} {054009} (\bibinfo {year} {2018})},\ \Eprint {https://arxiv.org/abs/1807.00718} {arXiv:1807.00718 [hep-ph]} \BibitemShut {NoStop}%
\bibitem [{\citenamefont {Lin}\ and\ \citenamefont {Zou}(2018)}]{Lin:2018nqd}%
  \BibitemOpen
  \bibfield  {author} {\bibinfo {author} {\bibfnamefont {Y.-H.}\ \bibnamefont {Lin}}\ and\ \bibinfo {author} {\bibfnamefont {B.-S.}\ \bibnamefont {Zou}},\ }\bibinfo {title} {{Hadronic molecular assignment for the newly observed $\Omega^*$ state}},\ \href {https://doi.org/10.1103/PhysRevD.98.056013} {\bibfield  {journal} {\bibinfo  {journal} {Phys. Rev. D}\ }\textbf {\bibinfo {volume} {98}},\ \bibinfo {pages} {056013} (\bibinfo {year} {2018})},\ \Eprint {https://arxiv.org/abs/1807.00997} {arXiv:1807.00997 [hep-ph]} \BibitemShut {NoStop}%
\bibitem [{\citenamefont {Pavao}\ and\ \citenamefont {Oset}(2018)}]{Pavao:2018xub}%
  \BibitemOpen
  \bibfield  {author} {\bibinfo {author} {\bibfnamefont {R.}~\bibnamefont {Pavao}}\ and\ \bibinfo {author} {\bibfnamefont {E.}~\bibnamefont {Oset}},\ }\bibinfo {title} {{Coupled channels dynamics in the generation of the $\Omega (2012)$ resonance}},\ \href {https://doi.org/10.1140/epjc/s10052-018-6329-4} {\bibfield  {journal} {\bibinfo  {journal} {Eur. Phys. J. C}\ }\textbf {\bibinfo {volume} {78}},\ \bibinfo {pages} {857} (\bibinfo {year} {2018})},\ \Eprint {https://arxiv.org/abs/1808.01950} {arXiv:1808.01950 [hep-ph]} \BibitemShut {NoStop}%
\bibitem [{\citenamefont {Huang}\ {\it et~al.}(2018)\citenamefont {Huang}, \citenamefont {Liu}, \citenamefont {Lu}, \citenamefont {Xie},\ and\ \citenamefont {Geng}}]{Huang:2018wth}%
  \BibitemOpen
  \bibfield  {author} {\bibinfo {author} {\bibfnamefont {Y.}~\bibnamefont {Huang}}, \bibinfo {author} {\bibfnamefont {M.-Z.}\ \bibnamefont {Liu}}, \bibinfo {author} {\bibfnamefont {J.-X.}\ \bibnamefont {Lu}}, \bibinfo {author} {\bibfnamefont {J.-J.}\ \bibnamefont {Xie}},\ and\ \bibinfo {author} {\bibfnamefont {L.-S.}\ \bibnamefont {Geng}},\ }\bibinfo {title} {{Strong decay modes $\bar{K}\Xi$ and $\bar{K}\Xi\pi$ of the $\Omega(2012)$ in the $\bar{K}\Xi(1530)$ and $\eta\Omega$ molecular scenario}},\ \href {https://doi.org/10.1103/PhysRevD.98.076012} {\bibfield  {journal} {\bibinfo  {journal} {Phys. Rev. D}\ }\textbf {\bibinfo {volume} {98}},\ \bibinfo {pages} {076012} (\bibinfo {year} {2018})},\ \Eprint {https://arxiv.org/abs/1807.06485} {arXiv:1807.06485 [hep-ph]} \BibitemShut {NoStop}%
\bibitem [{\citenamefont {Lu}\ {\it et~al.}(2020)\citenamefont {Lu}, \citenamefont {Zeng}, \citenamefont {Wang}, \citenamefont {Xie},\ and\ \citenamefont {Geng}}]{Lu:2020ste}%
  \BibitemOpen
  \bibfield  {author} {\bibinfo {author} {\bibfnamefont {J.-X.}\ \bibnamefont {Lu}}, \bibinfo {author} {\bibfnamefont {C.-H.}\ \bibnamefont {Zeng}}, \bibinfo {author} {\bibfnamefont {E.}~\bibnamefont {Wang}}, \bibinfo {author} {\bibfnamefont {J.-J.}\ \bibnamefont {Xie}},\ and\ \bibinfo {author} {\bibfnamefont {L.-S.}\ \bibnamefont {Geng}},\ }\bibinfo {title} {{Revisiting the $\Omega(2012)$ as a hadronic molecule and its strong decays}},\ \href {https://doi.org/10.1140/epjc/s10052-020-7944-4} {\bibfield  {journal} {\bibinfo  {journal} {Eur. Phys. J. C}\ }\textbf {\bibinfo {volume} {80}},\ \bibinfo {pages} {361} (\bibinfo {year} {2020})},\ \Eprint {https://arxiv.org/abs/2003.07588} {arXiv:2003.07588 [hep-ph]} \BibitemShut {NoStop}%
\bibitem [{\citenamefont {Ikeno}\ {\it et~al.}(2020)\citenamefont {Ikeno}, \citenamefont {Toledo},\ and\ \citenamefont {Oset}}]{Ikeno:2020vqv}%
  \BibitemOpen
  \bibfield  {author} {\bibinfo {author} {\bibfnamefont {N.}~\bibnamefont {Ikeno}}, \bibinfo {author} {\bibfnamefont {G.}~\bibnamefont {Toledo}},\ and\ \bibinfo {author} {\bibfnamefont {E.}~\bibnamefont {Oset}},\ }\bibinfo {title} {{Molecular picture for the $\Omega(2012)$ revisited}},\ \href {https://doi.org/10.1103/PhysRevD.101.094016} {\bibfield  {journal} {\bibinfo  {journal} {Phys. Rev. D}\ }\textbf {\bibinfo {volume} {101}},\ \bibinfo {pages} {094016} (\bibinfo {year} {2020})},\ \Eprint {https://arxiv.org/abs/2003.07580} {arXiv:2003.07580 [hep-ph]} \BibitemShut {NoStop}%
\bibitem [{\citenamefont {Ikeno}\ {\it et~al.}(2022)\citenamefont {Ikeno}, \citenamefont {Liang}, \citenamefont {Toledo},\ and\ \citenamefont {Oset}}]{Ikeno:2022jpe}%
  \BibitemOpen
  \bibfield  {author} {\bibinfo {author} {\bibfnamefont {N.}~\bibnamefont {Ikeno}}, \bibinfo {author} {\bibfnamefont {W.-H.}\ \bibnamefont {Liang}}, \bibinfo {author} {\bibfnamefont {G.}~\bibnamefont {Toledo}},\ and\ \bibinfo {author} {\bibfnamefont {E.}~\bibnamefont {Oset}},\ }\bibinfo {title} {{Interpretation of the $\Omega_c \to \pi^+ \Omega(2012)\to \pi^+ (K\Xi)$ relative to $\Omega_c \to \pi^+ K\Xi$ from the $\Omega(2012)$ molecular perspective}},\ \href {https://doi.org/10.1103/PhysRevD.106.034022} {\bibfield  {journal} {\bibinfo  {journal} {Phys. Rev. D}\ }\textbf {\bibinfo {volume} {106}},\ \bibinfo {pages} {034022} (\bibinfo {year} {2022})},\ \Eprint {https://arxiv.org/abs/2204.13396} {arXiv:2204.13396 [hep-ph]} \BibitemShut {NoStop}%
\bibitem [{\citenamefont {Liu}\ {\it et~al.}(2021)\citenamefont {Liu}, \citenamefont {Huang}, \citenamefont {Ping},\ and\ \citenamefont {Chen}}]{Liu:2020yen}%
  \BibitemOpen
  \bibfield  {author} {\bibinfo {author} {\bibfnamefont {X.}~\bibnamefont {Liu}}, \bibinfo {author} {\bibfnamefont {H.}~\bibnamefont {Huang}}, \bibinfo {author} {\bibfnamefont {J.}~\bibnamefont {Ping}},\ and\ \bibinfo {author} {\bibfnamefont {D.}~\bibnamefont {Chen}},\ }\bibinfo {title} {{Investigating $\Omega(2012)$ as a molecular state}},\ \href {https://doi.org/10.1103/PhysRevC.103.025202} {\bibfield  {journal} {\bibinfo  {journal} {Phys. Rev. C}\ }\textbf {\bibinfo {volume} {103}},\ \bibinfo {pages} {025202} (\bibinfo {year} {2021})},\ \Eprint {https://arxiv.org/abs/2010.15398} {arXiv:2010.15398 [hep-ph]} \BibitemShut {NoStop}%
\bibitem [{\citenamefont {Zeng}\ {\it et~al.}(2020)\citenamefont {Zeng}, \citenamefont {Lu}, \citenamefont {Wang}, \citenamefont {Xie},\ and\ \citenamefont {Geng}}]{Zeng:2020och}%
  \BibitemOpen
  \bibfield  {author} {\bibinfo {author} {\bibfnamefont {C.-H.}\ \bibnamefont {Zeng}}, \bibinfo {author} {\bibfnamefont {J.-X.}\ \bibnamefont {Lu}}, \bibinfo {author} {\bibfnamefont {E.}~\bibnamefont {Wang}}, \bibinfo {author} {\bibfnamefont {J.-J.}\ \bibnamefont {Xie}},\ and\ \bibinfo {author} {\bibfnamefont {L.-S.}\ \bibnamefont {Geng}},\ }\bibinfo {title} {{Theoretical study of the $\Omega(2012)$ state in the $\Omega_c^0 \to \pi^+ \Omega(2012)^- \to \pi^+ (\bar{K}\Xi)^-$ and $\pi^+ (\bar{K}\Xi\pi)^-$ decays}},\ \href {https://doi.org/10.1103/PhysRevD.102.076009} {\bibfield  {journal} {\bibinfo  {journal} {Phys. Rev. D}\ }\textbf {\bibinfo {volume} {102}},\ \bibinfo {pages} {076009} (\bibinfo {year} {2020})},\ \Eprint {https://arxiv.org/abs/2006.15547} {arXiv:2006.15547 [hep-ph]} \BibitemShut {NoStop}%
\bibitem [{\citenamefont {Xiao}\ and\ \citenamefont {Zhong}(2018)}]{Xiao:2018pwe}%
  \BibitemOpen
  \bibfield  {author} {\bibinfo {author} {\bibfnamefont {L.-Y.}\ \bibnamefont {Xiao}}\ and\ \bibinfo {author} {\bibfnamefont {X.-H.}\ \bibnamefont {Zhong}},\ }\bibinfo {title} {{Possible interpretation of the newly observed $\Omega(2012)$ state}},\ \href {https://doi.org/10.1103/PhysRevD.98.034004} {\bibfield  {journal} {\bibinfo  {journal} {Phys. Rev. D}\ }\textbf {\bibinfo {volume} {98}},\ \bibinfo {pages} {034004} (\bibinfo {year} {2018})},\ \Eprint {https://arxiv.org/abs/1805.11285} {arXiv:1805.11285 [hep-ph]} \BibitemShut {NoStop}%
\bibitem [{\citenamefont {Aliev}\ {\it et~al.}(2018{\natexlab{a}})\citenamefont {Aliev}, \citenamefont {Azizi}, \citenamefont {Sarac},\ and\ \citenamefont {Sundu}}]{Aliev:2018yjo}%
  \BibitemOpen
  \bibfield  {author} {\bibinfo {author} {\bibfnamefont {T.~M.}\ \bibnamefont {Aliev}}, \bibinfo {author} {\bibfnamefont {K.}~\bibnamefont {Azizi}}, \bibinfo {author} {\bibfnamefont {Y.}~\bibnamefont {Sarac}},\ and\ \bibinfo {author} {\bibfnamefont {H.}~\bibnamefont {Sundu}},\ }\bibinfo {title} {{Nature of the $\Omega (2012)$ through its strong decays}},\ \href {https://doi.org/10.1140/epjc/s10052-018-6375-y} {\bibfield  {journal} {\bibinfo  {journal} {Eur. Phys. J. C}\ }\textbf {\bibinfo {volume} {78}},\ \bibinfo {pages} {894} (\bibinfo {year} {2018}{\natexlab{a}})},\ \Eprint {https://arxiv.org/abs/1807.02145} {arXiv:1807.02145 [hep-ph]} \BibitemShut {NoStop}%
\bibitem [{\citenamefont {Aliev}\ {\it et~al.}(2018{\natexlab{b}})\citenamefont {Aliev}, \citenamefont {Azizi}, \citenamefont {Sarac},\ and\ \citenamefont {Sundu}}]{Aliev:2018syi}%
  \BibitemOpen
  \bibfield  {author} {\bibinfo {author} {\bibfnamefont {T.~M.}\ \bibnamefont {Aliev}}, \bibinfo {author} {\bibfnamefont {K.}~\bibnamefont {Azizi}}, \bibinfo {author} {\bibfnamefont {Y.}~\bibnamefont {Sarac}},\ and\ \bibinfo {author} {\bibfnamefont {H.}~\bibnamefont {Sundu}},\ }\bibinfo {title} {{Interpretation of the newly discovered $\Omega$(2012)}},\ \href {https://doi.org/10.1103/PhysRevD.98.014031} {\bibfield  {journal} {\bibinfo  {journal} {Phys. Rev. D}\ }\textbf {\bibinfo {volume} {98}},\ \bibinfo {pages} {014031} (\bibinfo {year} {2018}{\natexlab{b}})},\ \Eprint {https://arxiv.org/abs/1806.01626} {arXiv:1806.01626 [hep-ph]} \BibitemShut {NoStop}%
\bibitem [{\citenamefont {Polyakov}\ {\it et~al.}(2019)\citenamefont {Polyakov}, \citenamefont {Son}, \citenamefont {Sun},\ and\ \citenamefont {Tandogan}}]{Polyakov:2018mow}%
  \BibitemOpen
  \bibfield  {author} {\bibinfo {author} {\bibfnamefont {M.~V.}\ \bibnamefont {Polyakov}}, \bibinfo {author} {\bibfnamefont {H.-D.}\ \bibnamefont {Son}}, \bibinfo {author} {\bibfnamefont {B.-D.}\ \bibnamefont {Sun}},\ and\ \bibinfo {author} {\bibfnamefont {A.}~\bibnamefont {Tandogan}},\ }\bibinfo {title} {{$\Omega(2012)$ through the looking glass of flavour SU(3)}},\ \href {https://doi.org/10.1016/j.physletb.2019.03.054} {\bibfield  {journal} {\bibinfo  {journal} {Phys. Lett. B}\ }\textbf {\bibinfo {volume} {792}},\ \bibinfo {pages} {315} (\bibinfo {year} {2019})},\ \Eprint {https://arxiv.org/abs/1806.04427} {arXiv:1806.04427 [hep-ph]} \BibitemShut {NoStop}%
\bibitem [{\citenamefont {Liu}\ {\it et~al.}(2020)\citenamefont {Liu}, \citenamefont {Wang}, \citenamefont {L\"u},\ and\ \citenamefont {Zhong}}]{Liu:2019wdr}%
  \BibitemOpen
  \bibfield  {author} {\bibinfo {author} {\bibfnamefont {M.-S.}\ \bibnamefont {Liu}}, \bibinfo {author} {\bibfnamefont {K.-L.}\ \bibnamefont {Wang}}, \bibinfo {author} {\bibfnamefont {Q.-F.}\ \bibnamefont {L\"u}},\ and\ \bibinfo {author} {\bibfnamefont {X.-H.}\ \bibnamefont {Zhong}},\ }\bibinfo {title} {{$\Omega$ baryon spectrum and their decays in a constituent quark model}},\ \href {https://doi.org/10.1103/PhysRevD.101.016002} {\bibfield  {journal} {\bibinfo  {journal} {Phys. Rev. D}\ }\textbf {\bibinfo {volume} {101}},\ \bibinfo {pages} {016002} (\bibinfo {year} {2020})},\ \Eprint {https://arxiv.org/abs/1910.10322} {arXiv:1910.10322 [hep-ph]} \BibitemShut {NoStop}%
\bibitem [{\citenamefont {Arifi}\ {\it et~al.}(2022)\citenamefont {Arifi}, \citenamefont {Suenaga}, \citenamefont {Hosaka},\ and\ \citenamefont {Oh}}]{Arifi:2022ntc}%
  \BibitemOpen
  \bibfield  {author} {\bibinfo {author} {\bibfnamefont {A.~J.}\ \bibnamefont {Arifi}}, \bibinfo {author} {\bibfnamefont {D.}~\bibnamefont {Suenaga}}, \bibinfo {author} {\bibfnamefont {A.}~\bibnamefont {Hosaka}},\ and\ \bibinfo {author} {\bibfnamefont {Y.}~\bibnamefont {Oh}},\ }\bibinfo {title} {{Strong decays of multistrangeness baryon resonances in the quark model}},\ \href {https://doi.org/10.1103/PhysRevD.105.094006} {\bibfield  {journal} {\bibinfo  {journal} {Phys. Rev. D}\ }\textbf {\bibinfo {volume} {105}},\ \bibinfo {pages} {094006} (\bibinfo {year} {2022})},\ \Eprint {https://arxiv.org/abs/2201.10427} {arXiv:2201.10427 [hep-ph]} \BibitemShut {NoStop}%
\bibitem [{\citenamefont {Wang}\ {\it et~al.}(2023)\citenamefont {Wang}, \citenamefont {L\"u}, \citenamefont {Xie},\ and\ \citenamefont {Zhong}}]{Wang:2022zja}%
  \BibitemOpen
  \bibfield  {author} {\bibinfo {author} {\bibfnamefont {K.-L.}\ \bibnamefont {Wang}}, \bibinfo {author} {\bibfnamefont {Q.-F.}\ \bibnamefont {L\"u}}, \bibinfo {author} {\bibfnamefont {J.-J.}\ \bibnamefont {Xie}},\ and\ \bibinfo {author} {\bibfnamefont {X.-H.}\ \bibnamefont {Zhong}},\ }\bibinfo {title} {{Toward discovering the excited $\Omega$ baryons through nonleptonic weak decays of $\Omega_c$}},\ \href {https://doi.org/10.1103/PhysRevD.107.034015} {\bibfield  {journal} {\bibinfo  {journal} {Phys. Rev. D}\ }\textbf {\bibinfo {volume} {107}},\ \bibinfo {pages} {034015} (\bibinfo {year} {2023})},\ \Eprint {https://arxiv.org/abs/2203.04458} {arXiv:2203.04458 [hep-ph]} \BibitemShut {NoStop}%
\bibitem [{\citenamefont {Wang}\ {\it et~al.}(2007)\citenamefont {Wang}, \citenamefont {Huang}, \citenamefont {Zhang}, \citenamefont {Yu},\ and\ \citenamefont {Liu}}]{Wang:2007bf}%
  \BibitemOpen
  \bibfield  {author} {\bibinfo {author} {\bibfnamefont {W.-L.}\ \bibnamefont {Wang}}, \bibinfo {author} {\bibfnamefont {F.}~\bibnamefont {Huang}}, \bibinfo {author} {\bibfnamefont {Z.-Y.}\ \bibnamefont {Zhang}}, \bibinfo {author} {\bibfnamefont {Y.-W.}\ \bibnamefont {Yu}},\ and\ \bibinfo {author} {\bibfnamefont {F.}~\bibnamefont {Liu}},\ }\bibinfo {title} {{$\Omega \omega$ states in a chiral quark model}},\ \href {https://doi.org/10.1088/0253-6102/48/4/025} {\bibfield  {journal} {\bibinfo  {journal} {Commun. Theor. Phys.}\ }\textbf {\bibinfo {volume} {48}},\ \bibinfo {pages} {695} (\bibinfo {year} {2007})}\BibitemShut {NoStop}%
\bibitem [{\citenamefont {Ablikim}\ {\it et~al.}(2024)\citenamefont {Ablikim} {\it et~al.}}]{BESIII:2023mlv}%
  \BibitemOpen
  \bibfield  {author} {\bibinfo {author} {\bibfnamefont {M.}~\bibnamefont {Ablikim}} {\it et~al.} (\bibinfo {collaboration} {BESIII}),\ }\bibinfo {title} {{Study of excited $\Xi$ states in $\psi(3686) \to K^- \Lambda \Xi +c.c.$}},\ \href {https://doi.org/10.1103/PhysRevD.109.072008} {\bibfield  {journal} {\bibinfo  {journal} {Phys. Rev. D}\ }\textbf {\bibinfo {volume} {109}},\ \bibinfo {pages} {072008} (\bibinfo {year} {2024})},\ \Eprint {https://arxiv.org/abs/2308.15206} {arXiv:2308.15206 [hep-ex]} \BibitemShut {NoStop}%
\bibitem [{\citenamefont {Gutsche}\ and\ \citenamefont {Lyubovitskij}(2020)}]{Gutsche:2019eoh}%
  \BibitemOpen
  \bibfield  {author} {\bibinfo {author} {\bibfnamefont {T.}~\bibnamefont {Gutsche}}\ and\ \bibinfo {author} {\bibfnamefont {V.~E.}\ \bibnamefont {Lyubovitskij}},\ }\bibinfo {title} {{Strong decays of the hadronic molecule $\Omega^\ast (2012)$}},\ \href {https://doi.org/10.1088/1361-6471/abcb9f} {\bibfield  {journal} {\bibinfo  {journal} {J. Phys. G}\ }\textbf {\bibinfo {volume} {48}},\ \bibinfo {pages} {025001} (\bibinfo {year} {2020})},\ \Eprint {https://arxiv.org/abs/1912.10894} {arXiv:1912.10894 [hep-ph]} \BibitemShut {NoStop}%
\bibitem [{\citenamefont {Jia}\ {\it et~al.}(2019)\citenamefont {Jia} {\it et~al.}}]{Belle:2019zco}%
  \BibitemOpen
  \bibfield  {author} {\bibinfo {author} {\bibfnamefont {S.}~\bibnamefont {Jia}} {\it et~al.} (\bibinfo {collaboration} {Belle}),\ }\bibinfo {title} {{Search for $\Omega(2012)\to K\Xi(1530) \to K\pi\Xi$ at Belle}},\ \href {https://doi.org/10.1103/PhysRevD.100.032006} {\bibfield  {journal} {\bibinfo  {journal} {Phys. Rev. D}\ }\textbf {\bibinfo {volume} {100}},\ \bibinfo {pages} {032006} (\bibinfo {year} {2019})},\ \Eprint {https://arxiv.org/abs/1906.00194} {arXiv:1906.00194 [hep-ex]} \BibitemShut {NoStop}%
\bibitem [{\citenamefont {Jia}\ {\it et~al.}(2022)\citenamefont {Jia} {\it et~al.}}]{Belle:2022mrg}%
  \BibitemOpen
  \bibfield  {author} {\bibinfo {author} {\bibfnamefont {S.}~\bibnamefont {Jia}} {\it et~al.} (\bibinfo {collaboration} {Belle}),\ }\href@noop {} {\bibinfo {title} {{Observation of $\Omega(2012)^- \to \Xi(1530)\bar{K}$ and measurement of the effective couplings of $\Omega(2012)^-$ to $\Xi(1530)\bar{K}$ and $\Xi\bar{K}$}}} (\bibinfo {year} {2022}),\ \Eprint {https://arxiv.org/abs/2207.03090} {arXiv:2207.03090 [hep-ex]} \BibitemShut {NoStop}%
\bibitem [{\citenamefont {Navas}\ {\it et~al.}(2024)\citenamefont {Navas} {\it et~al.}}]{ParticleDataGroup:2024cfk}%
  \BibitemOpen
  \bibfield  {author} {\bibinfo {author} {\bibfnamefont {S.}~\bibnamefont {Navas}} {\it et~al.} (\bibinfo {collaboration} {Particle Data Group}),\ }\bibinfo {title} {{Review of particle physics}},\ \href {https://doi.org/10.1103/PhysRevD.110.030001} {\bibfield  {journal} {\bibinfo  {journal} {Phys. Rev. D}\ }\textbf {\bibinfo {volume} {110}},\ \bibinfo {pages} {030001} (\bibinfo {year} {2024})}\BibitemShut {NoStop}%
\bibitem [{\citenamefont {Molina}\ {\it et~al.}(2024)\citenamefont {Molina}, \citenamefont {Liang}, \citenamefont {Xiao}, \citenamefont {Sun},\ and\ \citenamefont {Oset}}]{Molina:2023uko}%
  \BibitemOpen
  \bibfield  {author} {\bibinfo {author} {\bibfnamefont {R.}~\bibnamefont {Molina}}, \bibinfo {author} {\bibfnamefont {W.-H.}\ \bibnamefont {Liang}}, \bibinfo {author} {\bibfnamefont {C.-W.}\ \bibnamefont {Xiao}}, \bibinfo {author} {\bibfnamefont {Z.-F.}\ \bibnamefont {Sun}},\ and\ \bibinfo {author} {\bibfnamefont {E.}~\bibnamefont {Oset}},\ }\bibinfo {title} {{Two states for the $\Xi(1820)$ resonance}},\ \href {https://doi.org/10.1016/j.physletb.2024.138872} {\bibfield  {journal} {\bibinfo  {journal} {Phys. Lett. B}\ }\textbf {\bibinfo {volume} {856}},\ \bibinfo {pages} {138872} (\bibinfo {year} {2024})},\ \Eprint {https://arxiv.org/abs/2309.03618} {arXiv:2309.03618 [hep-ph]} \BibitemShut {NoStop}%
\bibitem [{\citenamefont {Liang}\ {\it et~al.}(2024)\citenamefont {Liang}, \citenamefont {Molina},\ and\ \citenamefont {Oset}}]{Liang:2024fsv}%
  \BibitemOpen
  \bibfield  {author} {\bibinfo {author} {\bibfnamefont {W.-H.}\ \bibnamefont {Liang}}, \bibinfo {author} {\bibfnamefont {R.}~\bibnamefont {Molina}},\ and\ \bibinfo {author} {\bibfnamefont {E.}~\bibnamefont {Oset}},\ }\bibinfo {title} {{$\Omega_c \to \pi^+ \, (\pi^0,\, \eta)\, \pi \Xi^*,\,\pi^+ \, (\pi^0,\, \eta)\, \bar K \Sigma^*$ reactions and the two $\Xi(1820)$ states}},\ \href {https://doi.org/10.1103/PhysRevD.110.036005} {\bibfield  {journal} {\bibinfo  {journal} {Phys. Rev. D}\ }\textbf {\bibinfo {volume} {110}},\ \bibinfo {pages} {036005} (\bibinfo {year} {2024})},\ \Eprint {https://arxiv.org/abs/2404.18882} {arXiv:2404.18882 [hep-ph]} \BibitemShut {NoStop}%
\bibitem [{\citenamefont {Duan}\ {\it et~al.}(2024)\citenamefont {Duan}, \citenamefont {Song}, \citenamefont {Liang},\ and\ \citenamefont {Oset}}]{Duan:2024ygq}%
  \BibitemOpen
  \bibfield  {author} {\bibinfo {author} {\bibfnamefont {M.-Y.}\ \bibnamefont {Duan}}, \bibinfo {author} {\bibfnamefont {J.}~\bibnamefont {Song}}, \bibinfo {author} {\bibfnamefont {W.-H.}\ \bibnamefont {Liang}},\ and\ \bibinfo {author} {\bibfnamefont {E.}~\bibnamefont {Oset}},\ }\bibinfo {title} {{On the search for the two poles of the $\Xi (1820)$ in the $\psi (3686) \to \bar{\Xi }^+ \bar{K}^0 \Sigma ^{*-}(\pi ^- \Lambda )$ decay}},\ \href {https://doi.org/10.1140/epjc/s10052-024-13293-5} {\bibfield  {journal} {\bibinfo  {journal} {Eur. Phys. J. C}\ }\textbf {\bibinfo {volume} {84}},\ \bibinfo {pages} {947} (\bibinfo {year} {2024})},\ \Eprint {https://arxiv.org/abs/2405.03622} {arXiv:2405.03622 [hep-ph]} \BibitemShut {NoStop}%
\bibitem [{\citenamefont {Gamermann}\ {\it et~al.}(2010)\citenamefont {Gamermann}, \citenamefont {Nieves}, \citenamefont {Oset},\ and\ \citenamefont {Ruiz~Arriola}}]{Gamermann:2009uq}%
  \BibitemOpen
  \bibfield  {author} {\bibinfo {author} {\bibfnamefont {D.}~\bibnamefont {Gamermann}}, \bibinfo {author} {\bibfnamefont {J.}~\bibnamefont {Nieves}}, \bibinfo {author} {\bibfnamefont {E.}~\bibnamefont {Oset}},\ and\ \bibinfo {author} {\bibfnamefont {E.}~\bibnamefont {Ruiz~Arriola}},\ }\bibinfo {title} {{Couplings in coupled channels versus wave functions: application to the X(3872) resonance}},\ \href {https://doi.org/10.1103/PhysRevD.81.014029} {\bibfield  {journal} {\bibinfo  {journal} {Phys. Rev. D}\ }\textbf {\bibinfo {volume} {81}},\ \bibinfo {pages} {014029} (\bibinfo {year} {2010})},\ \Eprint {https://arxiv.org/abs/0911.4407} {arXiv:0911.4407 [hep-ph]} \BibitemShut {NoStop}%
\bibitem [{\citenamefont {Liang}\ and\ \citenamefont {Oset}(2020)}]{Liang:2020jtw}%
  \BibitemOpen
  \bibfield  {author} {\bibinfo {author} {\bibfnamefont {W.-H.}\ \bibnamefont {Liang}}\ and\ \bibinfo {author} {\bibfnamefont {E.}~\bibnamefont {Oset}},\ }\bibinfo {title} {{Testing the origin of the ``$f_1(1420)$" with the $\bar K p \to \Lambda (\Sigma) K\bar K \pi$ reaction}},\ \href {https://doi.org/10.1140/epjc/s10052-020-7966-y} {\bibfield  {journal} {\bibinfo  {journal} {Eur. Phys. J. C}\ }\textbf {\bibinfo {volume} {80}},\ \bibinfo {pages} {407} (\bibinfo {year} {2020})},\ \Eprint {https://arxiv.org/abs/2001.03299} {arXiv:2001.03299 [hep-ph]} \BibitemShut {NoStop}%
\bibitem [{\citenamefont {Debastiani}\ {\it et~al.}(2017)\citenamefont {Debastiani}, \citenamefont {Aceti}, \citenamefont {Liang},\ and\ \citenamefont {Oset}}]{Debastiani:2016xgg}%
  \BibitemOpen
  \bibfield  {author} {\bibinfo {author} {\bibfnamefont {V.~R.}\ \bibnamefont {Debastiani}}, \bibinfo {author} {\bibfnamefont {F.}~\bibnamefont {Aceti}}, \bibinfo {author} {\bibfnamefont {W.-H.}\ \bibnamefont {Liang}},\ and\ \bibinfo {author} {\bibfnamefont {E.}~\bibnamefont {Oset}},\ }\bibinfo {title} {{Revising the $f_1(1420)$ resonance}},\ \href {https://doi.org/10.1103/PhysRevD.95.034015} {\bibfield  {journal} {\bibinfo  {journal} {Phys. Rev. D}\ }\textbf {\bibinfo {volume} {95}},\ \bibinfo {pages} {034015} (\bibinfo {year} {2017})},\ \Eprint {https://arxiv.org/abs/1611.05383} {arXiv:1611.05383 [hep-ph]} \BibitemShut {NoStop}%
\bibitem [{\citenamefont {Lin}\ {\it et~al.}(2024)\citenamefont {Lin}, \citenamefont {Li}, \citenamefont {Liang}, \citenamefont {Chen},\ and\ \citenamefont {Oset}}]{Lin:2023ajb}%
  \BibitemOpen
  \bibfield  {author} {\bibinfo {author} {\bibfnamefont {J.-X.}\ \bibnamefont {Lin}}, \bibinfo {author} {\bibfnamefont {J.-T.}\ \bibnamefont {Li}}, \bibinfo {author} {\bibfnamefont {W.-H.}\ \bibnamefont {Liang}}, \bibinfo {author} {\bibfnamefont {H.-X.}\ \bibnamefont {Chen}},\ and\ \bibinfo {author} {\bibfnamefont {E.}~\bibnamefont {Oset}},\ }\bibinfo {title} {{$J/\psi$ decays into $\omega (\phi) f_1(1285)$ and $\omega (\phi) ``f_1(1240)"$}},\ \href {https://doi.org/10.1140/epjc/s10052-024-12405-5} {\bibfield  {journal} {\bibinfo  {journal} {Eur. Phys. J. C}\ }\textbf {\bibinfo {volume} {84}},\ \bibinfo {pages} {52} (\bibinfo {year} {2024})},\ \Eprint {https://arxiv.org/abs/2310.19213} {arXiv:2310.19213 [hep-ph]} \BibitemShut {NoStop}%
\end{thebibliography}%
\end{document}